# Fare-Free Bus Service and CO2 Reductions: Evidence from a Natural Experiment

by

Anna Alberini, Javier Bas (corresponding author), and Cinzia Cirillo[1]

16 September 2025

**Abstract:** We devise a difference-in-difference study design to assess the impact of fare-free bus service in Alexandria, located in the Washington, DC metro area. Our surveys show modest to no effect, with at most 6% more residents in Alexandria increasing their bus usage compared to control locations. We find no effect on ground-level ozone or road crashes, suggesting little to no impact on road traffic.

One-third of respondents in control locations indicated they would use buses more frequently if fare-free service were available in their areas. Based on the respondent-reported reductions in car miles, the program led to a reduction of 0.294 to 0.494 tons of CO2 per year, or 5% to 9% of the average annual emissions from a US car, at a cost of \$70-\$120 per ton of CO2. We predict a CO2 reduction of 0.454 tons per year, equivalent to 8% of the average US car's annual emissions if the fare-free bus covered all of the study areas.

**JEL classification:** R41, R42, R48, Q54.

**Keywords:** public transportation; fare-free bus service; ridership; $CO_2$ emissions; Difference-in-difference.

[1] Anna Alberini is a professor in AREC at the University of Maryland, College Park, and an affiliate of the Maryland Transportation Institute. Email: aalberin@umd.edu. Address: 2210 Symons Hall 7998 Regents Drive College Park, Maryland, U.S. 20742-5535. Javier Bas (corresponding author) is an associate professor at the Universidad Autónoma de Madrid. Email: javier.bas@uam.es. Address: Facultad de CC. Económicas y Empresariales, C/ Francisco Tomás y Valiente, 5, 28049 Madrid, España. Cinzia Cirillo is a professor in the Department of Civil and Transportation Engineering at the University of Maryland, College Park, and the interim director of the Maryland Transportation Institute. Email: ccirillo@umd.edu. Address: 3250 Jeong H. Kim Engineering Building, 7998 Regents Drive College Park, Maryland, U.S. 20742-5535. We are grateful to Bavan Nadarajah, Pragna Yalamanchili and Rin Futara for outstanding research assistance, and to Dan Goldfarb and Martin Barna for their support and helpful suggestions. This research was supported by NSF RAPID award 2153689 entitled "Fare Free Public Transportation - A Full Scale Natural Experiment in Alexandria, Virginia." Javier Bas's work was supported by grant PID2022-139614NB-C21 funded by MCIN/AEI /10.13039/501100011033/FEDER, EU.

## 1. Introduction

In the United States, the transportation sector is responsible for some 29% of the emissions of carbon dioxide ($CO_2$) and other greenhouse gases (US EPA, 2024). Emissions from the transportation sector have declined by some 6% compared to their 2005 levels (Congressional Budget Office, 2022), but—in sharp contrast with the electricity generation sector—the trend remains relatively flat. The European Union is experiencing similar difficulties in their efforts to reduce greenhouse gas emissions from transportation (European Environment Agency, 2024).

In addition to considering alternative fuels (such as electricity or hydrogen), fuel pricing and incentive policies, governments and international organizations recommend shifting as much travel as possible to public transportation, which can save "up to 2.2 tons of carbon emissions annually per individual" (https://www.un.org/en/actnow/transport). Yet, in the US public transit accounts for a very small share of all trips and has suffered a decline in ridership starting with the Covid-19 pandemic (US DOT, 2024). Much of this decline is likely to be permanent, as people continue to work from home and resort more and more frequently to online shopping (US DOT, 2024). Would people use public transit more, and avoid emissions from private cars, if public transportation were free?

On the 5[th] of September 2021, rides became free on the DASH buses in Alexandria, a city on the Potomac River just south of Washington, DC, in the DC metro area. Fare collection was eliminated, with the support of the Alexandria City Council, in an effort to "promote bus ridership" and "alleviate a financial burden for low-income riders still recovering from the adverse economic impact of the Covid-19 pandemic." [2] The City of Alexandria subsequently received funding from the Virginia Department of Rail and Public Transit to support the fare-free program for the next three years. Alexandria's free-riding program followed others implemented elsewhere in the US, city-wide or on a small scale, in other countries such as Estonia (Cats et al., 2014, 2017) or Luxembourg (Carr and Hesse, 2020), and/or nominal fares that effectively make public transportation near-free at the regional level as in Spain (Ensor, 2023) or Germany (Rozynek, 2024).

---

[2] See https://www.dashbus.com/free/.

Information about the fare-free transit in Alexandria is posted on DASH's website, and on billboards around the area, including in Washington, DC.  At the beginning of August 2023, DASH reported that ridership on its buses had grown to the point of reaching an all-time high, attributing this achievement to the fare-free program.[3] This finding appears to be based on the comparison between recent and historical DASH ridership figures.

Ascertaining the effect of a free bus program, however, is complicated by the existence of a number of confounding factors, including the reopening of the economy after the early phases of the Covid-19 pandemic, transitory and permanent changes in school and work schedules and hence in commuting, and the large increase in motor fuel prices over 2021 and 2022. All of these factors may have contributed to changing travel patterns, regardless of the availability of free bus service. For instance, He et al. (2022) highlight the significant impact of the Covid-19 pandemic on essential transit riders, showing how travel behaviors were altered due to the pandemic's constraints and changes in work habits. Similarly, Hu and Chen (2021) discuss the socioeconomic disparities in the decline of transit ridership during the pandemic, pointing to a nuanced landscape of travel behavior that complicates the assessment of fare-free programs. Additionally, Kutela et al. (2022) provide insights into the long-term effects of Covid-19 on transportation facilities, suggesting that some changes in travel behavior may persist beyond the pandemic. Furthermore, the trade-offs between greenhouse gas emissions and travel time, as explored by Aziz and Ukkusuri (2014), and the potential emission reductions from a shift towards public transport, as measured by Carroll et al. (2019), indicate that factors such as environmental concerns and efficiency in travel time also influence travel choices. Lastly, O'Riordan et al. (2022) examine the broader question of mobility demand and emissions from passenger transport, which underscores the complexity of interpreting changes in travel patterns in the context of various overlapping influences.

---

[3] See https://www.alxnow.com/2023/08/01/alexandrias-dash-bus-network-sees-record-high-ridership-after-going-fare-free/?mc_cid=7407e54616&mc_eid=9410b1ebde.

In this paper, we examine the impact of fare-free bus rides on people's travel habits, using a rigorous study design that seeks to account for the abovementioned factors. We ask three research questions. First, do people use the bus more often as a result of free bus service? Second, has the program left a "trail" of effects on road traffic and environmental quality? Third, how many private car miles travelled (VMT) were avoided by the free bus program—and would be avoided *if* the program were adopted on a larger scale—and what is the extent and cost of the associated reduction in $CO_2$ emissions?

We interpret the fare-free program as a natural experiment, and devise a difference-in-difference study design that relies on a "treatment group" (commuter, residents, road crashes or air pollution levels in Alexandria) and on a "control group" (the same, but in Washington, DC, and the "outer ring" of counties around Washington, DC, in the DC metro area), and on data before and after the fare-free program was implemented. The control group allows us to "filter out" the effect of the factors that have affected everyone regardless of the program, such as the reopening of the economy, changed schedules, trends in gasoline prices, etc.

We use data from federal and local governments to examine the effect of the free bus program on the use of the bus for commuting purposes. Since this analysis misses the effect—if any—of the free bus program on people that do not work—retirees, homemakers, students, and others—we collect data about bus rides and trips by other modes through a survey of residents of the treatment and control locations, which we administered in the early summer of 2022. We find that commuters do not seem to have been affected by the free bus program; the survey of residents indicates at best a modest increase in the use of the bus attributable to the free bus program. But this differential effect (5.80%) is barely statistically significant at the 5% level, and not robust with respect to different outcome variables (intensive v. extensive margins), focusing on certain groups of people (e.g., workers) or controlling for additional covariates. Ridership figures from WMATA indicate that Metrobus ridership in Alexandria—not just DASH ridership—has increased since the pandemic and has in many cases reached in 2023 all-time highs. Data limitations prevent us from examining the effect of the free bus program on traffic volumes, but we do have data on road crashes and local air pollution, and they too are not significantly impacted by the fare free program.

It is important to find out whether fare-free programs reduce private car VMT and hence the associated greenhouse gas emissions—or would if they were adopted on a larger scale. We use the respondents' own estimates of the VMT reduced and the emissions rate of the average car in the US (404 g/mile) to compute that the fare-free program avoided 0.494 emissions tons of $CO_2$ for the average car in Alexandria—approximately 9% of the annual emissions of the average US car—at a forgone revenue cost of \$70/metric ton.[4] These estimates become only slightly less favorable (0.3 tons of $CO_2$ per year, or \$120/metric ton) when we match the vehicle owned by the respondent with its EPA-assessed fuel economy and use the latter to compute each car's emissions of CO2.

---

[4] The US EPA reports that, at 404 g $CO_2$ per mile, the average passenger car in the US emits a total of 4.6 metric tons of $CO_2$ a year (https://www.epa.gov/greenvehicles/greenhouse-gas-emissions-typical-passenger-vehicle).

This paper contributes to the nascent literature that seeks to apply experimental or quasi-experimental methods to assess the effect of fare-free (or almost fare-free) public transportation (Bull et al., 2021; Ortega et al., 2023). While fare free transit programs (FFTPs) have existed at various locations for decades,[5] they have received much recent attention, as they have been considered and adopted at some locales as a form of income transfer to help support the population during the Covid-19 pandemic.

Earlier assessments for the US, based on 39 cases, have concluded that limited fare-free programs have resulted in 20-60% increase in ridership (Volinski, 2012). Only 5-10% of the new transit trips, however, displaced private car trips. Only few of the actual fare-free transit programs remained in place long enough to observe their long-term effects on travel behavior, and those that did were primarily in Europe. The evidence from the latter is mixed: At some locations (e.g., Hasselt, Belgium; Verachtert, 2013; Aubagnac, France, and Chapel Hill, NC; Volinski, 2012); Örebro and Kristinhamn, Sweden) readership increased quickly, requiring investment in the fleet and expansion of the network, and eventually forcing the local authorities to reintroduce fares. Other locations were able to accommodate the increase in ridership without reinstituting fares, but the increase in ridership was due mainly due to more frequent trips by existing public transport users (Ramboll, 2013). Other locations yet (e.g., Tallinn, Estonia) were able to achieve significant improvements in the mobility of low-income residents (Cats et al., 2014, 2017).

---

[5] Most of the services that operate fare-free are limited in scale (e.g. airport shuttle; university campus bus), reserved to selected categories of users (e.g. tourists, students, elderly), and/or temporary (e.g. during big events/disasters; campaigns to reduce car use). Kębłowski (2020) shows that the vast majority of fare-free public transport schemes worldwide are small-scale, targeted, or short-lived.

Mobility, however, is not the only outcome of interest when examining FFTPs (Börjesson and Eliasson, 2025). Fearnley (2013) reviews several studies, and notes the possibility of substitution from walking and cycling into riding the bus, which results in road safety gains. Keblowski (2000) focuses on FFTPs that are temporary, spatially limited, or restricted to certain groups of users, discussing implications for revenue (and hence the financing of the service) and sustainability. Liang and Wang (2025) deploy a difference-in-difference design to assess the effect of a FFTP on air quality in Fuzhou, China. Their strategy is to define non-working days (weekends and holidays) as the control group, as the FFTP is in place only on workdays. They document that PM2.5 fell by 2.1% compared to the control group days; the associated benefits are six times as large as the forgone fare. Gregersen et al. (2024) evaluate the introduction of free public transport for residents using a difference-in-differences design based on automatic passenger counts and find a 7.5% increase in overall boardings and a 9.1% increase during morning peak hours, although the latter is not statistically significant. Vieira et al. (2025) use data from 11 different household surveys and a regression discontinuity design to assess the effect of a FFTP on the elderly in a developing country with substantial transit use—Brazil. Their strategy is to compare the use of public transportation among people just below and just above the minimum age required to qualify for free transit. They find that free transit rides increases ridership by 9.4% but shorten trips by 8.2%, and has no effect on car usage.

## 2. Background

### 2.1 Study Area

We focus on residents of the city of Alexandria (our treatment location), plus Washington, DC, and the "outer ring" of counties and independent cities around it, namely Arlington County, Fairfax County, Prince William County, Falls Church, Manassas and Manassas Park in Virginia, and Montgomery, Prince George's, Calvert, Charles and Frederick counties in Maryland (our control locations; see figure 1).[6]

Attention is restricted to this area for three reasons. First, this area contains a mix of high-density as well as suburban and relatively rural locations. Second, the presence of the federal government and other large employers ensures complex commuting patterns that include travel into Washington, DC (about one-third of workers), as well as to workplaces in the outer ring (58-61% of workers, according to the 2019 and 2022 State of the Commute Surveys). The 2019 and 2022 State of the Commute surveys further indicate that the one-way commute distance is stable over across these two waves (median 14 miles, mean about 17 miles) (WashCog, 2020, 2023). Third, the area is covered by Metrobus and Metrorail service, for a total of 99.7 million riders a year (WMATA, 2023). Metrobus and Metrorail are well integrated with county and local buses, and DASH is an example of the latter.

According to the 2019 American Community Survey, the median household income in our study area is $100,000; the first quartile of the distribution of household income is $50,000, and the third $165,000. Educational attainment is high, with a 50.47% of adults over 25 holding a bachelor's or higher degree, as reported by the U.S. Census Bureau.

*2.2 Bus Use and Ridership in the DC Metro Area*

The most recent information about individual bus use before the pandemic comes from the Regional Travel Survey (RTS) conducted by the Metropolitan Washington Council of Government in 2017-18,[7] which asks respondents to report all trips made within 24 hours of the day of the survey. The trip rosters of the residents of our study area indicate that about three-quarters of the trips are made by car and that the travel mode is a bus only in just over 2% of the trips. Further inspection of the data reveals that a bus is used in 2813 out of 75,897 trips by adults, and that the bus is the main mode of transportation in 1816 of these 2813 trips—about two-thirds of them.

---

[6] We exclude from our analysis Loudoun County in Virginia.
[7] See https://www.mwcog.org/documents/2022/04/22/regional-travel-survey-in-depth-analysis-featured-publications-regional-travel-survey/.

Buses take people home or to work (about 40% each), shopping (about 10% of the trips) and to school (about 2% of the trips), accounting for some 5% of "direct commutes" (home-work and work-home trips). Own cars are the most common travel mode in our study area. Alexandria residents however appear to be somewhat less likely to drive by themselves or with others, and more likely to walk or take the subway—but just as likely to take the bus as their counterparts in the other study locations.[8] When attention is restricted to "direct commutes," Alexandria residents are slightly more likely to use the bus as the main travel mode than workers living at the other locations (6.94% v. 5.08%), and are definitely more likely to use the bus together with other modes (19.19% v. 8.90%).

*2.3 Fare-free Bus Service*

The residents of the DC metro area are no strangers to the notion of free bus ridership. On March 24, 2020, at the beginning of the Covid-19 pandemic, WMATA stopped collecting the fare on its Metrobuses. Passengers were to board the bus through the backdoor to avoid exposing the drivers to the virus; since it was not feasible to relocate the fareboxes to the back of the buses, the agency simply had to forgo the fare. This was, of course, only one of the major disruptions to public transit service experienced during the Covid-19 pandemic and the associated lockdowns. Ridership fell due to massive service cuts, and the facts that the use of public transit was discouraged, many employers (including the federal government) resorted to telecommuting and school instruction moved to online. WMATA resumed regular fare collection ($2) on Jan. 3, 2021.

---

[8] Specifically, Alexandria residents drive alone in 60.96% of the trips, and with others in 7.77%. They walk in 15.54% of the trips, take Metrorail in 9.59% of them, and ride a bus in 2.48%. The percentages in the other study locations are, in order, 68.95%, 9.07%, 9.89%, 6.24%, and 2.48%.

WMATA posts ridership figures on its website.[9] Figure 2 displays systemwide ridership for Metrobuses and Metrorail (the subway system in the DC metro area) from January 2018 to July 2023. Both types of transit experienced a negative shock in ridership during the early phases of the Covid-19 pandemic, and have been recovering since, but have not yet reached their pre-pandemic ridership levels. Metrorail appears to have recovered faster than Metrobus.

DASH adopted the same fare policy as WMATA during the pandemic, and likewise resumed charging fare on Jan. 3, 2021. On Sept. 5, 2021, rides became free on the DASH buses in the city of Alexandria. These events are summarized in figure 3. At the time of this writing, DASH operates a total of 11 routes, including two "Old Town circulators" and the King Street Trolley. Most of its routes start and/or end at Metrorail stations.

## 3. What are the Effects of Fare Free Transit?

In theory, if public transit is free, people should use it more often, perhaps replacing the use of their car or other private motorized transportation to go to work, shopping and to run errands. DASH indeed reports a dramatic increase in ridership since the inception of the program—and so does WMATA, which, after a nine-month hiatus in 2020, resumed collecting and currently still collects fares from its passengers, for many routes close to or overlapping with the DASH routes. (See figures A.1 and A.2 in the Appendix.) One problem when examining these figures is that ridership is typically measured as boardings, boardings are tallied when passengers swipe their SmartTrip card, and DASH riders no longer do that because the bus is free.[10] Another difficulty is that DASH buses are free to everyone, without distinguishing between residents and visitors, but in this paper we are primarily interested in the effect of the free bus program on residents.

---

[9] See https://www.wmata.com/initiatives/ridership-portal/.

We look for evidence of increased bus use in the commuting behavior data collected by the federal and the local government. Since commuter surveys miss the travel mode choice decisions of persons that do not work—such as students, retirees, homemakers and others—we conduct our own survey of residents of the City of Alexandria and of the control locations to inquire about their use of buses and whether it has changed since the implementation of the DASH FFTP.

If a FFTP truly gets people to ride transit more often and forgo private motorized transportation, and this effect is sufficiently pronounced, it might be possible to find a "trail" of it in traffic volumes, ambient concentrations of air pollutants typically associated with vehicle tailpipe emissions, and in road crashes. Earlier studies (e.g., Fernley, 2013) have suggested that if free public transit attracts primarily pedestrians and cyclists, removing them from the road may result in fewer accidents, or at least fewer crashes that involve pedestrians and cyclists.

Of course, it is also perfectly possible that a free bus program simply increases welfare for those that take the bus without inducing any additional ridership. Finally, some have wondered whether FFTPs might even *discourage* ridership, if people fear that they end up making buses too crowded, less clean, or perhaps less safe.

---

[10] DASH bus drivers are instructed to manually count passengers, but a common complaint is that this interferes with the safe operation of the bus (Alexandria Transit Company, 2022).

## 4. Data

*4.1 Commuting Behavior*

We use data from up to up to three waves of the American Community Survey (2019-2022, skipping 2020), conducted by the federal government, and two waves of the State of the Commute (2019 and 2022), administered by the Metropolitan Washington Council of Governments, to examine commuting trips in our study area.

The American Community Survey (ACS) collects extensive information about sociodemographics, labor force participation and work status, including the mode of transportation used to commute to the workplace and the usual duration of the commute in minutes. We use the data from the 2019, 2021 and 2022 waves of the American Community Survey. Residents of Alexandria account for 990, 934, and 986 of the workers documented therein in 2019, 2021, and 2022, respectively, versus 28,369, 28,723 and 27,491 residents of the control locations. The shares of ACS commuters who use the bus to go to work are displayed in tables 1 and 2.

The State of the Commute Survey is conducted by the Metropolitan Washington Council of Governments and focuses exclusively on workers, who are to indicate how many times a week they report to a workplace outside of the home or work from home, and how they reach that workplace. Respondents must identify their travel mode to the workplace (or the fact that they telecommute from home) for each of the 5 weekdays. Weights are provided in the individual-level dataset.

The 2019 SOC contains a total of 693 workers who live in Alexandria and 6844 that live in our control locations.  The 2022 SOC covers 793 workers who live in Alexandria and 6950 that live in the control locations. Importantly, the online questionnaire allows respondents to pick only one mode for each of the 5 workweek days. If someone chooses the bus, we interpret that to mean that the bus was the main or most important among multiple modes that may also include, for example, walking or driving a car to and from a "park-and-ride" lot. We collapse the responses to the Monday through Friday commute mode question in the 2019 and 2022 State of the Commute Surveys into broad categories, as shown in table 3.

*4.2 Our Survey of Residents of the Area*

Our survey questionnaire elicits information about the number of motor vehicles available to the respondent's household, and the specifics of the vehicle the respondent uses most often, including make, model, model-year, fuel economy and miles driven in the last 12 months.

The next section inquires about private and public transportation. How often did the respondent use their own motor vehicles, the bus,[11] Metrorail, taxis or Uber, or a bicycle in the last month? How often did they walk to their destinations? We offered five possible response options ranging from "daily" to "never." We then asked the respondent the same questions, but for August 2021, namely just before the free bus policy went into effect in Alexandria.

We asked respondents whether they were aware of fare discounts or the availability of free surface or underground public transportation where they live. Residents of Alexandria were then informed that bus rides had been free of charge since September 2021 in their city, and asked to indicate if and how they had changed their transportation patterns as a result. All other respondents, who live in areas where the buses are not free, were asked what they would do *if* bus rides became free. Would they take the bus more often? Would they drive their cars less? If so, approximately how many private car miles would they give up per week?[12]

---

[11] We asked about "the bus" without distinguishing for the type of service and/or the company or government agency that provided it, as it was important for us to avoid conveying the idea that the questionnaire was part of a market research survey associated with specific entities.

The questionnaire was programmed in Qualtrics and self-administered online by a total of 997 respondents recruited from the Qualtrics panel in Alexandria (N=338), Washington, DC, and our control counties (N=659) between June 24, and July 5, 2022 (see Table A.1).[13] We imposed the requirement that the sample be comprised of an even number of men and women, and be stratified for income, with 25% of the respondents in each quartile of the distribution of household income in the area.[14] [15]

Summary statistics about the sociodemographics of the survey participants are displayed in tables A.2-A.5 in the Appendix. Compared to the control locations, our Alexandria sample boasts a larger share of the full-time or part-time employed respondents (88% v. 66%). The educational attainment of the respondent is however similar across Alexandria and the control locations, with t tests indicating that the shares of schooling level are not statistically different. This is clearly a highly educated sample, with more than half of the respondents in each group having a four-year college degree, a master's degree or a PhD, or a professional degree. This level of education is similar to that of the population of the area. Both the treatment and control group mirror the area's household income quartiles. The treatment and control locations are also similar in terms of ethnicity (see table A.5 in the Appendix).[16]

---

[12] It is worth noting that WMATA offers monthly passes valid for bus and rail (https://www.wmata.com/fares/Monthly-Pass/), but it remains unclear how prevalent their use was at the time of the survey. Riders using monthly passes would behave as if the bus were free, since the marginal price of an extra bus trip is effectively zero in both cases. This information was not collected in the survey.

[13] Since Frederick County never discontinued the free ride policy it adopted at the beginning of the pandemic, and thus experienced no change in the "free bus rides" status, we consider it part of true control group. For good measure, in the next section we repeat all of the analyses after dropping the 44 residents of Frederick County. The results are always virtually identical.

[14] We remind the reader that at our study locations the first quartile of household income was, at the time of the survey, $55,000; the second was $100,000; and the third $165,000.

[15] Qualtrics report that "good completes" (i.e., valid completed questionnaires) account for 8.97% of the contacts they attempted. About 81.80% did not meet the screening questions (e.g., they were not 18 or older, or they lived outside of the study area), 8.18% were over the quotas, and 1.05% were classified as speeders or otherwise failed data quality checks.

Table 4 reports the percentages of the respondents who use the bus daily, 3-5 times a week, 1-2 times a week, occasionally, and never. The table shows that Alexandria residents are and were more likely to take the bus than residents at the control locations—both before and almost a year after the implementation of the free bus policy.[17]

Our questionnaire checks the respondents' knowledge of free bus service in the area where they live. Some 20% of the respondents believe that anyone in the area where they live is entitled to free bus service. Unsurprisingly, this share is higher among Alexandria respondents, but not by an overwhelming margin (26% v. 17%). It is noteworthy that almost three-quarters of the Alexandria respondents do not appear to be aware of the free bus service available to them.

*4.3 Additional Data*

Although the DC, MD and VA Departments of Transportation (DoTs) maintain a number of traffic flow monitors in our study area, traffic volumes are available for Alexandria only for 2019, and only at a location (I-95, a major interstate) that is unlikely to mirror local traffic. This prevents us from examining whether traffic volumes have changed differentially in Alexandria since the inception of the program.

---

[16] Some 17.15% of the respondents consider themselves Hispanic. The composition of the sample by race is displayed in table A.5. White/Caucasians account for over 60% of the sample, African Americans for some 20%, and Asians for just under 8%. Native American and Hawaiian/Pacific Islander account for just under 4%.
[17] A probit regression based on the most recent reported bus usage (Table A.6 in the Appendix) suggests that this tendency does not disappear when we account for income, work status, length of the commute, education level and car ownership. The only variables that are significantly associated with riding the bus at least occasionally are whether the respondent works full- or part-time work (positively associated) and whether they own a car (negatively associated).

We have hourly ozone measurements (in parts per million) from one air quality monitor in Alexandria and a handful of others at the control locations every day from March to October (the ozone season) from 2015 to 2022.[18] Figure 4 plots the monthly average of the hourly data at the Alexandria and the control location monitors, showing that they track each other well, both before and after the beginning of the free bus program, and confirming the seasonal nature of ozone pollution.

Road traffic accidents data are provided by the Virginia, Maryland and DC Departments of Transportation (DoTs) for the period from January 2015 to June 2022. We have the exact location and time of the crash, whether pedestrians or cyclists were involved, whether injuries or fatalities occurred, and other administrative information. Summary information about the crash data is displayed in table 5.

## 5. Methods

We rely on the difference-in-difference study design, as we have observations from treated and control subjects (or entities) before and after the treatment in Alexandria and at the control locations (Card and Krueger, 1994; Cook and Campbell, 1979). The treatment is the DASH FFTP, which is "as good as randomly assigned" to anyone in the DASH service territory.[19]

With the ACS and State of the Commute surveys, which are cross-sections repeated twice, we estimate the average treatment effect on the treated (ATT) as $(\bar{y}_2^T - \bar{y}_2^C) - (\bar{y}_1^T - \bar{y}_1^C)$, where $\bar{y}$ is the share of residents who take the bus to commute to work, superscripts T and C denote the treatment and control groups, respectively, and subscripts 1 and 2 refer to before and after the FFTP has started.

---

[18] Other conventional air pollutants that are potentially sensitive to traffic include nitrogen dioxide and carbon monoxide, but the monitoring network maintained by the US EPA does not collect data for these pollutants in the city of Alexandria.

[19] That doesn't necessarily mean that our treatment group subjects or entities availed themselves of it. This is thus an "intention to treat" type of natural experiment (Hollis and Campbell, 1999; Angrist and Pischke, 2009).

Our own survey gathers information about residents' bus usage frequency before and after the free bus program was in place. We form a longitudinal dataset where each respondent contributes T=2 observations. Let $y$ denote the outcome of interest and D a dummy indicator denoting whether the free bus program is in effect at the location of individual $i$. The model is

(1) $$y_{it} = \alpha_i + \beta \cdot D_{it} + \gamma \cdot P_t + \varepsilon_{it},$$

where $i$ denotes the respondent, $t=1$ denotes August 2021, $t=2$ denotes the time of the survey (June or July 2022), and $P$ is the "post" dummy, i.e., a dummy that takes on a value of one if $t=2$ and zero otherwise. Note that dummy $D_{it}$ is equal to the treatment group dummy times the "post" dummy. Equation (1) includes an individual-specific fixed effect and deploys $P_t$ to capture the effect of events that affected everyone in the study area (e.g., the reopening of the economy after the pandemic, etc.). $\beta$ is the ATT of the free bus program.

Implicit in the study design and estimator are the assumptions that no one relocated from the control locations to Alexandria because of Alexandria's free bus program, and that the chance that residents of the control locations use the Alexandria buses is negligible. Also implicit is assumption of common trends across treatment and control units, which cannot be tested, not even partially, because T is equal to 2.

On taking first differences, we obtain:

(2) $$\Delta y_i = \beta \cdot \Delta D_i + \gamma + \Delta \varepsilon_i.$$

It should be noted that $\Delta D_i$—the change in treatment status—is the same as a dummy denoting whether the respondent lives in Alexandria. An unbiased estimate of $\beta$ is obtained by running

OLS on this regression, or, equivalently, by computing the average $\Delta y$ for the treatment group and subtracting the average $\Delta y$ of the control group (the difference-in-difference [DID] estimator).

This procedure is easily applied when $y$ is a continuous variable (for example, bus miles traveled), but a bit less intuitive in a situation like ours, as our questionnaire collects categorical data about bus use frequency. We thus define our outcome variable $y$ so that $\Delta y$ measures, in alternate specifications, the overall, an "extensive margin" and an "intensive margin," respectively, change in bus usage. By extensive margin, we mean whether someone never used the bus in the pre-policy period but does now, at least occasionally. By intensive margin, we mean that someone who used the bus previously has increased their bus usage since the beginning of the free bus program, switching, for example, from occasional use to once or twice a week. This implies that when we run our regressions, the usable sample must be redefined appropriately. In the extensive margin regressions, the usable sample includes only those respondents who did not use the bus in August 2021. In the intensive margin regressions, the usable sample includes only those individuals that used the bus before the beginning of the free bus program.

With the accident and pollution concentration data, we form panel datasets where each location or pollution monitor contributes T>2 observations to the panel. This time we have several time periods—not just one—before the implementation of the FFTP and so we must check the underlying assumption of common trends. With these data, we fit variants on

(3) $\qquad y_{it} = \alpha_i + \beta \cdot (Alex_i \times Post_t) + \tau_t + \mathbf{z_t}\boldsymbol{\gamma} + \varepsilon_{it}$

where y is an outcome of interest at location (or monitor) i in period t, Alex is a dummy denoting the treatment location (Alexandria) and Post is a dummy denoting that the free bus program is in force. $\tau_t$ is a time fixed effect, and **z** includes time-varying variables thought to be associated with the outcome $y$.[20]

    We note here that we do not expect the FFTP to have a measurable effect on ozone, and that we examine it in this paper merely for the sake of completeness: Ozone is a regional pollutant that affects a much broader area than our study locations and occurs under specific atmospheric and solar radiation conditions, and the City of Alexandria and the DASH service territory are likely too small to make a dent on the NOx and hydrocarbons emissions that eventually become tropospheric ozone—even if the FFTP had produced a dramatic reduction in private vehicle driving.

## 6. Results

### 6.1. Evidence from the Commuter Surveys

    Tables 1 and 2 report the shares of the workers in the 2019-2022 waves of the American Community Survey who use the bus to commute to work. Before the pandemic, these shares were around 5%; they fell by 40-50% after 2020, implying that, based on the simple "difference-in-difference" calculation reported in table 1, the increase in bus ridership among commuters attributable to the free bus program is less than 1%.[21] This effect is statistically insignificant, whether we compare 2019 with 2021 or 2022. Based on the ACS alone, one would reach the

---

[20] One such variant is a Poisson regression, which is appropriate with very low counts, such as those of crashes involving pedestrians and/or cyclist fatalities. The probability of observing exactly $y_{it}$ such crashes at location i in period t is $\Pr(y_{it}) = \lambda_{it}^{y_{it}} \cdot e^{\lambda_{it}}/y_{it}!$, where $\lambda_{it} = \exp{(\alpha_i + \tau_t + \beta \cdot (Alex_i \times Post_t))}$ is both the expectation and the variance of $y_{it}$. This Poisson model is estimated by the method of maximum likelihood. A further generalization of the Poisson is the negative binomial model, which, among other things, admits overdispersion (i.e., the variance of $y_{it}$ is higher than its expected value).

[21] Based on 2022 versus 2019, the effect attributable to the program is -0.0171-(-0.0221)=0.005, while its counterpart for 2021 against 2019 is -0.0203-(-0.02810=0.0078. As explained in Appendix A, assuming no other difference between the city of Alexandria and the control locations, this should be interpreted at the increase in bus ridership among Alexandria commuters once the decline in bus usage common to both Alexandria and the other locations has been accounted for.

conclusion that the availability of free bus rides has had virtually no effect on the use of the bus for commuting to work.[22]

One arrives at similar conclusions using the State of the Commute Surveys: The percentages displayed in table 3 allow us to calculate, for example, the effect of the free bus program on the likelihood of never taking the bus at all to go to work. This is equal to (0.9636-0.8844)-(0.9615-0.9273)=0.045. The standard error around this estimate is 0.0146. In other words, the free bus program appears to have *increased* the likelihood that a worker *never* takes the bus at all to go to work and this effect is statistically significant at the 1% level.

Similar calculations with the shares of workers that use the bus 3-5 times a week result in an estimated effect of -0.0281 (standard error 0.0127): Taking the bus to go to work 3-5 times a week has *fallen* by more in Alexandria than in the control locations. The effect is marginally statistically significant at the 5% level. The same calculations for those who report commuting by bus 1-2 times a week result in an estimated net effect of -0.0169 (standard error 0.0077), again revealing that bus commutes with this frequency have fallen in Alexandria to a slightly greater extent than in the control locations.

### 6.2. Evidence from Our Own Survey

Perusal of table 4 suggests only very small changes in the shares of the participants in our survey who take the bus with any frequency in the year since the beginning of the policy. At the control locations, the number of respondents who "never" take the bus has fallen by about 5%, and that of persons that take the bus 1-2 times a week has fallen by 1.4%. These persons appear to have moved to the "occasional" and "more than 1-2 times a week" categories, for a very small impact on each of these categories. Among the residents of Alexandria, there has been a small decline in the shares that take the bus "never," only "occasionally" or 1-2 times a week, and a small increase in the shares of daily (1.2%) and 3-5 times a week (3.2%) users. These increases are not individually statistically significant at the conventional levels.[23]

---

[22] It should be kept in mind, however, that only in 2022 was the fare-free program fully in place. In 2021, it was in place for only four months (September to the end of the year).

Table 6 reports several possible definitions of $\Delta y$ and the associated descriptive statistics. Table 7 displays the results from fitting OLS to equation (2), where $\Delta y$ is defined as a strict increase in the frequency of bus usage in col. (1) (e.g., going from "never" to "occasionally" or from "occasionally" to "daily"), the extensive margin (going from "never" to at least occasionally) in col. (2), and, iii) in cols. (3) and (4), two possible interpretations of the intensive margin.

Row (1), column (1) of table 7 shows that bus travel appears to have grown 5.80% faster in Alexandria, the free bus location, than elsewhere. This figure should be interpreted as meaning that the share of residents that have used the bus more often than they did in August 2021 has increased by 5.80% more in Alexandria than at the other locations. This estimate of the ATT is marginally significant at the 5% level. Likewise, column (2) shows that the share of those who previously *never* travelled by bus but do now, at least occasionally, has grown by 6.52% more at the free bus location. This effect is however barely significant at the 10% level. We emphasize that all ATTs capture the effect of the availability of fare-free DASH rides on the frequency with which our respondents took *any* bus—whether or not it was a DASH bus. We estimate a strict intensive margin ATT of 4.23% attributable to the availability of the free bus program, but this ATT is statistically insignificant at the conventional levels (column (3)). A broader definition of intensive margin, namely whether respondents have increased or maintained the frequency of their bus rides, results in a virtually zero ATT (column (4)).

---

[23] The questionnaire incorporated some redundancies that allow us to check the quality of the data. For example, residents of Alexandria were asked if they had changed their travel habits since the beginning of the free bus program. We checked the survey responses of the 108 persons who said they had been riding the bus more often since. For almost 80% of them the reported frequency of bus use in the previous month is indeed greater or equal to that in August 2021. This and other checks suggest to us that the survey responses are internally consistent.

Just in case residents of neighboring counties ride DASH buses, in which case they should be reassigned to the treatment group instead of being regarded as part of the control group, in the second row of table 6 we present the results of the same regressions, but with the sample limited to Alexandria, Washington, DC and Maryland counties residents, on the reasoning that DC and Maryland are unlikely to utilize DASH buses. The results are virtually unchanged, even though the t statistics for the Alexandria dummy coefficient are weaker in cols. (2)-(4) due to the smaller sample sizes. Attention is restricted to respondents who work full- or part-time in row (3) of table 6, finding that the ATTs are weaker—both in magnitude and statistical significance—than for the entire sample(s).[24]

Table 8 displays the results of similar regressions that further control for household income. The results are virtually the same as their counterparts in row (1) of table 7. When we enter further socioeconomic characteristics of the respondents, however, as we do in the regression of table A.7 in the Appendix, the coefficients on the Alexandria dummies become statistically insignificant at the conventional levels.

### 6.3 Accidents and Pollutant Ambient Concentrations

We tallied the number of road crashes reported to the DC, MD and VA DoTs in each month from January 2015 to June 2022, producing separate counts for all types of crashes, crashes involving pedestrians or cyclists (or both), and crashes that resulted in a pedestrian or cyclist fatality. The statistics shown in table 5 suggest a linear panel data model for the total number of crashes, and negative binomial models with location and time fixed effects when we examine pedestrian- or cyclist-involving crashes. Pedestrian and cyclist fatalities are, fortunately, even rarer and for this reason we fit a Poisson model with location fixed effects and time fixed effects to their monthly counts. Results are reported in table 9.

---

[24] We wondered whether the lack of statistical significance was due to the use of the linear probability model and associated robust standard errors and t statistics, but when we fit a probit model (which is estimated by the method of maximum likelihood and thus presumably produces the most efficient estimates) the ATTs and their p values are similar to those of the linear probability model (results available from the authors).

The ATT of the FFTP for all road crashes is positive and very imprecisely estimated, whereas it is in most cases negative when attention is restricted to pedestrians and cyclists. Even in these cases, the ATT of the FFTP is significant, and at no better than the 10% level, only when the full sample period is considered and pedestrians are examined, alone or together with cyclists.[25]

Table 10 reports the results of our ozone regressions. We fit equation (3) to the hourly ozone data at each monitor as well as to monthly average of each hourly measurement at each monitor.[26] All specifications include dummies for the day of the week, a major holiday dummy, monitor and time fixed effects.[27]

The estimation results in table 10 suggest that no meaningful changes in ozone levels since the inception of the FFTP. This is hardly surprising: Ozone is a regional pollutant that affects a much broader area than the DC metro area and forms at specific atmospheric and solar radiation conditions, and the City of Alexandria and the DASH service territory are presumably too small to make a dent on the NOx and hydrocarbon emissions that eventually become tropospheric ozone—even if the FFTP had produced a dramatic reduction in private vehicle driving.

---

[25] In those cases, the magnitude of the coefficient on the Alexandria × Post interaction implies a 30-35% decline in the expected number of crashes in the treatment area compared to the control locations.

[26] Equation (3) follows from a difference-in-difference study design, which means that we must check that treatment and control locations followed a common trend prior to the administration of the treatment. Using the monthly averages and a specification similar to (3), but limited to the data before the beginning of the FFTP, we find no statistically significant difference between the trend in Alexandria and that in the control location. (The coefficient on the time trend interacted with the Alexandria dummy is 2.86E-06 and its t statistic only 0.13.) With the hourly data, we find common trends only when the sample is restricted to 2018 and the later years, and for this reason we restrict the sample to 2018 and subsequent years when we fit equation (3) to the hourly data.

[27] We omit weather information such as temperature and precipitation, which usually contribute to pollution levels, because they are too similar across locations; they are thus subsumed into the time fixed effects.

*6.4 CO2 Emissions Reductions*

We would likewise expect modest CO2 emissions reduction from DASH's FFTP. Since CO2 emissions and the concentration of CO2 in the atmosphere do not get measured by regular air pollution monitors, we use an alternate approach to compute them.

In our survey, we asked respondents to assess how much driving they had given up since the free bus program—if they were residents of Alexandria—or how much driving they would give up if bus rides were free in their area. To assist them while they were seeking to answer this question, we instructed them to think about where they would go and how often, and what the distance to such destinations would be. The time frame they were to focus on was one week, and they were to consider the round-trip miles.

When asked if they had given up some car travel since the bus had become fare-free, 43.65% of the Alexandria residents said they had. By contrast, of the respondents from the control locations, 36.37% said that they would take the bus more often if buses were fare-free, and 69.88% said that they would give up some car travel.

We were able to elicit estimates of car VMT avoided from 736 respondents. The responses summarized in table 11 are based on a clean sample that excludes car VMT saved greater than 2049 miles (an obvious outlier). The average is about 31 miles per week, with only a negligible difference between the mean figures for Alexandria and control locations residents. It is however interesting that, as shown in figure 5, Alexandria residents are both more likely to report zero avoided VMT and to report relatively large avoided VMT figures than the residents of the control locations.

Using information from the 2022 American Community Survey, we constructed weights to make our sample comparable to the population in terms of income, age, gender and area of residence. The population-weighted average car VMT avoided per week by an Alexandria resident with a car is 23.52, which implies—assuming that the average car emits 404 g $CO_2$/mile[28]—0.494 metric tons of $CO_2$ emissions avoided per year. This figure can be multiplied by 1.59 (the average number of cars owned by an Alexandria household, according to official statistics) and 70,289 (the number of households in Alexandria) to arrive at annual emissions reductions of 56,120 metric tons from Alexandria residents.[29]

Since our survey respondents reported the make, model and model-year of the car they use, we obtained exact average miles per gallon (MPG) from www.fueleconomy.gov, and combined it with the content of CO2 per gallon[30] to arrive at a vehicle-specific emissions rate per mile. This alternate calculation results in 0.294 tons per population-representative respondent per year and 33,218 tons of CO2 avoided per year by Alexandria residents.

When the entire sample is used (i.e., the car miles respondents *say* they would avoid *if* the bus were free in their area, and the car miles saved as reported by the Alexandria residents), we arrive at a population-representative weekly average of 22.83 miles, and, assuming again that the average car emits 404 g $CO_2$/mile about 0.480 tons of $CO_2$ saved per year. Taking into account the number of households in the study area and the average number of cars per household, this would result in over two million tons of $CO_2$ avoided per year. The calculation based on exact information about the vehicle owned by the respondent produces similar figures—0.454 tons of CO2 per person and car, and over two million tons of $CO_2$ reduced per year area-wide.

---

[28] See https://www.epa.gov/greenvehicles/greenhouse-gas-emissions-typical-passenger-vehicle.
[29] This calculation takes the respondents' estimates of car VMT avoided at face value; it further assumes that they are not replaced in part or entirely by VMT driven by other household members.
[30] https://www.epa.gov/greenvehicles/greenhouse-gas-emissions-typical-passenger-vehicle.

## 7. Discussion and Conclusions

We have taken advantage of the establishment of fare-free transit in Alexandria, VA, which we interpret as a natural experiment, to estimate the effect of the availability of free rides on the residents' use of the bus. We have used a difference-in-difference study design, examining the *change* in bus use frequency among Alexandria residents *compared* to that among the residents of a broad area within the DC metro region where fares continued to be collected as usual. We have used a similar study design to see if the FFTP has left a "trail" in outcomes typically associated with road traffic and transit, such as road traffic accidents and conventional air pollution. This approach posits that there were no other concurrent policies or programs (e.g., other income transfer programs or extensive infrastructure projects) specific to Alexandria in the same period that could have affected travel behavior, road traffic, and air pollution patterns.

Using data from the American Community Survey and Washcog's State of the Commute just before the pandemic and in 2021 and 2022, we find that after the pandemic workers are less likely than before to use the bus for commuting purposes, and even *less* likely in Alexandria than in our control locations in the DC metro area. Using the data from our own survey of residents of the area (a group that includes retirees, homemakers, students, and others in addition to persons in the labor force), we find a modest effect of the fare-free program on the frequency of use of the bus. Almost 6% more residents of Alexandria experienced an increase in their frequency of bus rides than their counterparts in our control locations. The effect is barely statistically significant at the 5% level and is not robust to restricting the sample to certain groups (e.g., workers) or controlling for a number of covariates.

While we are unable to examine the effect of the FFTP on traffic volumes, due to lack of suitable data, we do check whether crashes and ozone levels have changed with the FTTP. The ATT of the free bus program does have a negative sign when attention is restricted to crashes involving pedestrians and/or cyclists, consistent with the notion that pedestrians and cyclists are less at risk when they substitute walking or cycling with free bus rides (Fearnley, 2013). But the effects are imprecisely estimated and are statistically significant, and only at the 10% level, in only two regressions out of 12.

As expected, the FFTP has not affected ozone levels either. This confirms Albalate et al.'s (2024) point, based on their evaluation of Spain's nationwide 2022 policy, which shows no detectable improvement in air quality, suggesting that fare-free or discounted transit alone may be insufficient to meaningfully reduce private car use and associated emissions without complementary measures, such as congestion pricing or parking restrictions. Similarly, Webster (2024) shows that Colorado's statewide 'Zero Fare for Better Air'—which offered free public transit during August 2022 with the explicit goal of reducing ozone pollution—failed to achieve detectable reductions in ozone levels, despite the 15-20% gains in transit usage.

If we trust our survey participants' reports of VMT avoided since the inception of the FFTP, the free bus program has some effect on private car tailpipe $CO_2$ emissions. Our assessments suggest that the average Alexandria resident cut down on VMT enough to reduce annual $CO_2$ emissions by 5% (according to a calculation that takes into account the specific make and model of the respondents' vehicle) to 9% (if we assume the emissions rate of the average car in the US fleet) compared to the average US car. The overall emissions reductions are modest, given the Alexandria's population and car fleet, but very cost-effective: DASH estimates that going fare-free implies a forgone revenue of up to $4 million in 2022 (Alexandria Transit Company, 2022), which, once divided by the $CO_2$ emissions reductions according to our two alternate calculations, results in cost figures of $120/metric ton $CO_2$ and $70/metric ton of $CO_2$, respectively. These figures are very favorable when compared with, for example, alternate initiatives in the US and elsewhere meant to reduce greenhouse gas emissions by promoting the adoption of electric vehicles. The latter's cost has generally been above $400/ton of $CO_2$ in the US (Xing et al., 2021; Sheldon et al., 2023) and has even reached €1000-2400 in Europe (in Germany: Haan et al., 2025; Alberini and Vance, 2025).

A larger scale program, i.e., one that covers our entire study area, would reduce $CO_2$ emissions by some two million tons—again, *if* we trust the respondents' announcements about what they would do if the free bus program were extended to their area, and rule out "latent" VMT by other members of the family or of the general public (in response to the increased availability of a family vehicle or the reduction in traffic on roads). These calculations assume that the forgone private car VMT are not replaced at all, or are replaced by the same VMT on other vehicles with zero tailpipe emissions, like electric cars or fully electric buses.[31]

If there is little evidence that residents—whether commuters or otherwise—ride the bus more often because it is free, what explains the record ridership figures reported by DASH? One possibility that comes to mind is that they might be due to visitors. Alexandria is a favorite place to stay for airline crews, businesspeople and other visitors to Washington DC, due to the availability and cost of accommodation, and the fact that it is so well connected to the capital. It is certainly convenient for these people to hop on a DASH bus to their final destinations or to reach a Metrorail station. But since anyone can board the DASH buses, we are unable to confirm or disprove this conjecture. Meanwhile, DASH reports that ridership continues to rise, with boarding records repeatedly broken in fiscal years 2023 (4.54 million), 2024 (5.34 million), and 546,000 boardings in a single month in 2025, with projections that in 2025 annual boardings on track will exceed 5.3 million (https://www.dashbus.com/dash-earns-2025-outstanding-community-program-award/).

---

[31] As of late 2024, out of the 101 active revenue buses in DASH's fleet, 14 are 100% electric, 47 are hybrid electric, and 42 are clean diesel. DASH added two electric buses in June 2025.

While our findings provide valuable insights into the effects of Alexandria's FFTP, several limitations must be acknowledged. First, our analysis relies partly on self-reported survey data, which may be affected by recall errors. Second, we ask people to estimate the reduction in car miles driven since the inception of the program (in Alexandria) or if a FFTP program were adopted in their area. It is perfectly possible that people may have driven fewer miles (or may drive fewer miles) even if they do not use the bus themselves, but we have no way of validating these estimates.

Third, data limitations prevent us from assessing the program's effects on traffic volumes, and our estimates of air quality rely on a single monitoring station in Alexandria, potentially missing localized variations. Fourth, we used a difference-in-difference design assumes parallel but potential spillover effects—such as control-area residents benefiting indirectly from DASH's fare-free rides—could attenuate the measured treatment effect. We have done our best to restrict the analysis to people that we deem unlikely to be affected by such possible spillovers (e.g., residents of DC or Maryland, which would be unlikely to use the DASH buses), but we cannot entirely exclude them.

Fifth, the specific socioeconomic and transport characteristics of Alexandria, combined with the strong integration of DASH and WMATA services, may limit the generalizability of our results to other urban settings. Finally, the study focuses on short-term impacts in a period that could still be affected by COVID-19-related disruptions; further research with longer-term data will be needed to fully understand the behavioral and environmental consequences of fare-free transit policies.

**Table 1.** American Community Survey: Share of workers who use the bus use for commuting and Difference-in-difference estimate of the ATT. 2019 v. 2022.

| Commute by bus[1] | Alexandria | Control locations |
|---|---|---|
| Before (2019) | 0.0564 | 0.0487 |
| After (2022) | 0.0393 | 0.0266 |
| Difference | -0.0171 | -0.0221 |
| **DID ATT** | 0.005 | |
| **se(ATT)** | 0.010097 | |

[1] weighted using ACS person weights.

**Table 2.** American Community Survey: Share of workers who use the bus use for commuting and Difference-in-difference estimate of the ATT. 2019 v. 2021.

| Commute by bus[1] | Alexandria | Control locations |
|---|---|---|
| Before (2019) | 0.0564 | 0.0487 |
| After (2021) | 0.0361 | 0.0206 |
| Difference | -0.0203 | -0.0281 |
| **DID ATT** | 0.0078 | |
| **se(ATT)** | 0.010359 | |

[1] weight using ACS person weights.

**Table 3.** State Of the Commute responses: Bus use to commute to work, 2019 and 2022 waves.

| State of the Commute wave | Location | 3-5x/week | 1-2x/week | Never |
|---|---|---|---|---|
| 2019 | Alexandria | 8.88% | 2.67% | 88.44% |
| | Control locations | 6.00% | 1.27% | 92.73% |
| 2022 | Alexandria | 2.34% | 1.29% | 96.36% |
| | Control locations | 2.27% | 1.58% | 96.15% |

**Table 4.** Own survey of DC metro area residents. Bus use frequency before and with the free bus program. Percentage of the treatment and control groups.

| Location | | Daily | 3-5 times a week | 1-2 times a week | Occasionally | Never |
|---|---|---|---|---|---|---|
| Alexandria | Before | 6.8% | 14.2% | 16.9% | 26.9% | 35.2% |
| | After | 8.0% | 17.5% | 16.0% | 23.7% | 34.9% |
| Control | Before | 5.0% | 5.9% | 10.7% | 16.4% | 62.0% |
| | After | 5.2% | 8.8% | 9.3% | 20.0% | 56.7% |

**Table 5.** Summary statistics about the number of road crashes. Average number of crashes per month, Jan 2015-Jun 2022.

| | Alexandria | Control locations |
|---|---|---|
| All | 120.33 | 590.43 |
| Involving pedestrians | 5.1 | 12.52 |
| Involving cyclists | 1.21 | 5.93 |
| Involving pedestrians or cyclists | 6.31 | 18.32 |
| With a pedestrian or cyclist fatality | 0.20 | 0.38 |

**Table 6.** Own survey of DC metro area residents. Possible definitions of $\Delta y$ and percentages in the treatment and control groups.

| Variable | Definition | Alexandria | Control locations | All |
|---|---|---|---|---|
| Morethan1 | Respondent is using the bus more frequently than in August 2021 | 22.19% | 16.39% | 18.36% |
| Morethan2 | Respondent did not use the bus in August 2021 but does now, at least occasionally | 21.85% | 15.32% | 16.79% |
| Morethan3 | Respondent did use the bus in August 2021, and now uses it more frequently | 22.37% | 18.14% | 20.13% |
| Morethan4 | Respondent did use the bus in August 2021, and now uses it just as or more frequently | 73.06% | 72.98% | 73.02% |

**Table 7.** Own survey of DC metro area residents: Coefficients on the Alexandria dummy in linear probability models of dependent variables morethanX. The model also includes a constant (omitted from this table). Robust standard errors in parentheses.

| sample | (1) morethan1 | (2) morethan2 | (3) morethan3 | (4) morethan4 |
|---|---|---|---|---|
| | (use of bus has strictly increase since last year) | (extensive margin: never used the bus last year, but now at least occasionally) | (intensive margin: used the bus at least occasionally last year, has strictly increased use since) | (intensive margin: used the bus at least occas. last year, has increased or stayed the same since) |
| (1) All locations | 0.0580** (0.0268) | 0.0652 (0.0419) | 0.0423 (0.0373) | 0.00075 (0.0412) |
| (2) Alexandria, Washington DC and Maryland locations only | 0.0675** (0.0289) | 0.0659 (0.0443) | 0.0503 (0.0404) | 0.0080 (0.0455) |
| (3) Only persons who work Full-time or part-time | 0.0432 (0.0306) | 0.0249 (0.0459) | 0.0404 (0.0430) | -0.0018 (0.0444) |

*: significant at the 10% level; **: significant at the 5% level; ***: significant at the 1% level.

**Table 8.** Own survey of DC metro area residents. Linear probability models of dependent variables morethanX. OLS estimates; robust standard errors in parentheses.

|  | (1) morethan1 | (2) morethan2 | (3) morethan3 | (4) morethan4 |
|---|---|---|---|---|
| Cons | 0.1281*** | 0.0739*** | 0.1842*** | 0.7059*** |
|  | (0.0246) | (0.0320) | (0.0282) | (0.0402) |
| Alexandria | 0.0582** | 0.0670* | 0.0424 | -0.0006 |
|  | (0.0268) | (0.0416) | (0.0373) | (0.0613) |
| Household income (thou. $) | 0.00032* | 0.00070* | -0.000026 | 0.00023 |
|  | (0.00019) | (0.00026) | (0.00027) | (0.00027) |
| N | 997 | 530 | 467 | 467 |

\*: significant at the 10% level; \*\*: significant at the 5% level; \*\*\*: significant at the 1% level.

Table 9. Accident data. Coefficient on Alexandria dummy × Post in regressions of the number of accidents in each month on Alexandria dummy × Post, location fixed effects, month fixed effects, and year fixed effects.

| Accident type | Type of model | Jan 2015-Jun 2022 | Jan 2018-Jun 2022 | Jan 2018-Jun 2022, excl. 2020 |
|---|---|---|---|---|
| All accidents reported to DC, MD and VA DoTs | Linear FE Model | 33.24 (47.49) | 32.70 (52.54) | 58.94 (49.06) |
| Accidents involving pedestrians reported to DC, MD and VA DoTs | Negative Binomial FE Model | -0.4267* (0.2308) | -0.2813 (0.2389) | -0.3442 (0.2476) |
| Accidents involving cyclists reported to DC, MD and VA DoTs | Negative Binomial FE Model | -0.1223 (0.3387) | 0.0556 (0.3500) | 0.0090 (0.3602) |
| Accidents involving pedestrians or cyclists reported to DC, MD and VA DoTs | Negative Binomial FE Model | -0.3540* (0.2162) | -0.1976 (0.2227) | -0.2641 (0.2319) |
| Accidents with pedestrian or cyclist fatalities reported to DC, MD and VA DoTs | Poisson FE Model | -0.2206 (0.6419) | -0.1246 (0.1699) | -0.1709 (0.1725) |

\*: significant at the 10% level; \*\*: significant at the 5% level; \*\*\*: significant at the 1% level.

Table 10. Effect of FFTP on ozone levels. Standard errors in parentheses clustered at the month-by-year level.

| | Monthly averages of the hourly data | | | Hourly data | |
|---|---|---|---|---|---|
| | Mar 2015-Jun 2022 | Mar 2018-Jun 2022 | Mar 2018-Jun 2022, excl. 2020 | Mar 2018-Jun 2022 | Mar 2018-Jun 2022, excl. 2020 |
| $\beta$ (ATT of free bus program on ozone levels) | 0.00014 (0.0009) | 0.00018 (0.00095) | 0.00033 (0.0010) | -4.26E-06 (0.0026) | 0.0000175 (0.002737) |
| Monitor fixed effects | Y | Y | Y | Y | Y |
| Time fixed effects | Month Year | Month Year | Month year | Month-by-year | Month-by-year |
| Holiday & day of the week dummies | Y | Y | Y | Y | Y |

*: significant at the 10% level; **: significant at the 5% level; ***: significant at the 1% level.

Table 11. Own survey of DC metro area residents. Car VMT per week reduced since the free bus/if the bus was free.

| | Mean | 10th percentile | 25th percentile | 50th percentile | 75th percentile | 90th percentile | 95th percentile |
|---|---|---|---|---|---|---|---|
| Control locations | 31.95 | 0 | 0 | 10 | 25 | 100 | 200 |
| Alexandria | 30.26 | 0 | 0 | 0 | 10 | 50 | 100 |
| Entire sample | 31.18 | 0 | 0 | 2 | 20 | 75 | 150 |

Based on 731 obs, having dropped obs with ≥2049 miles

**Figure 1.** Study Area: The "Outer Ring" of counties in the Washington, DC, metro area. Alexandria is the treatment location and the other counties/cities are the control locations.

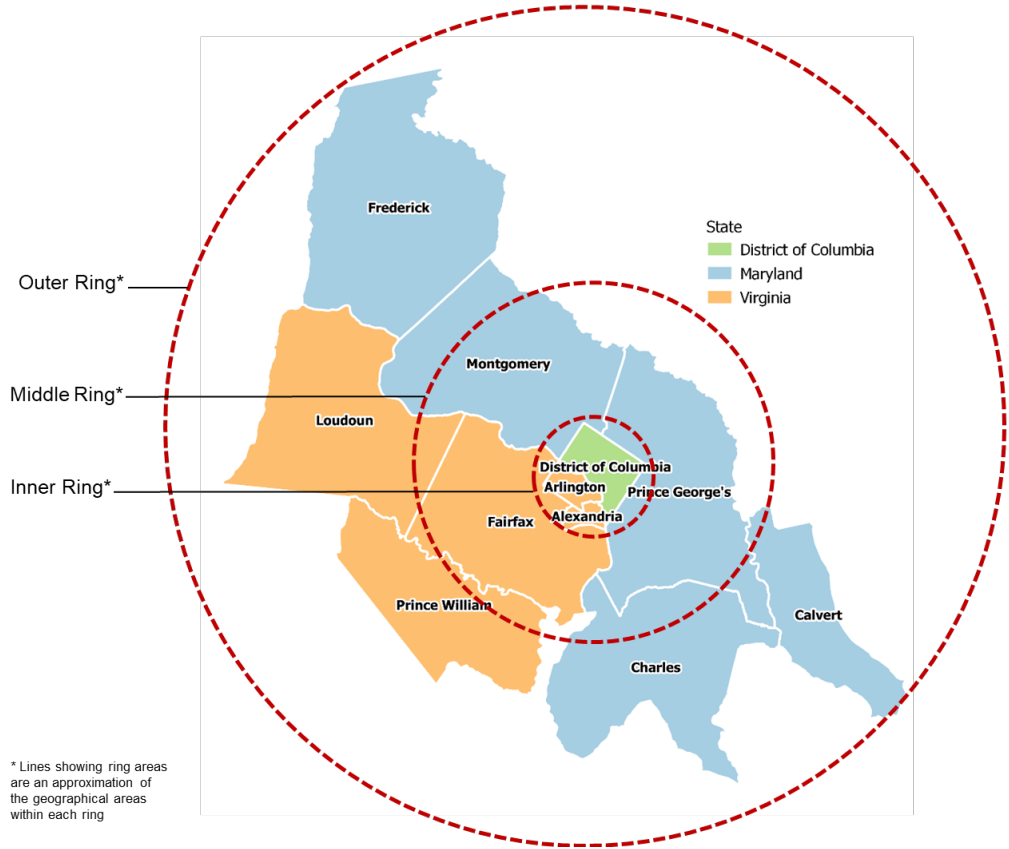

Source: Map created by Guangchen Zhao following a similar map in the 2022 State of the Commute (WashCog, 2023)

**Figure 2.** Average daily entries for Metrobus and Metrorail, Jan 2018 to ~~Jul~~ Apr 2023. Source: WMATA.

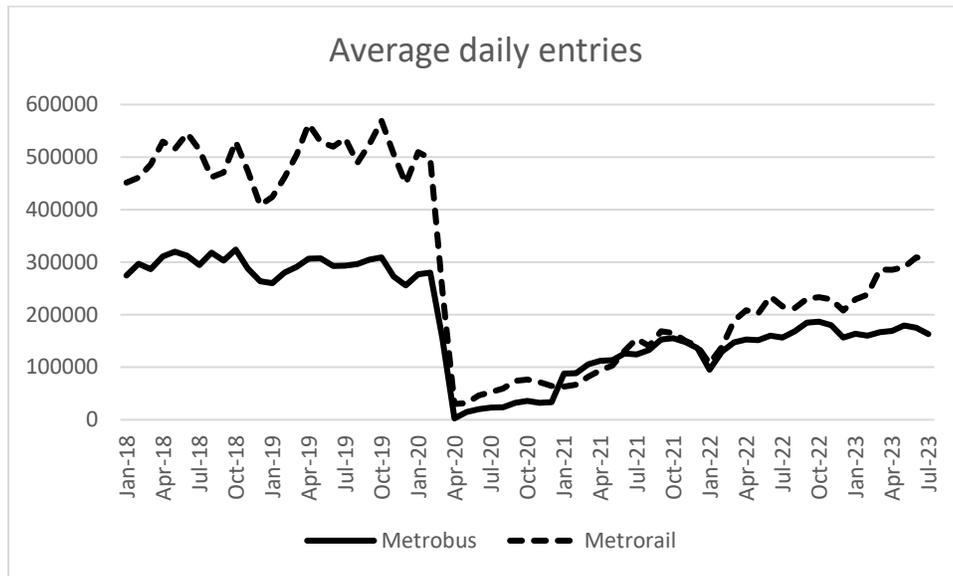

**Figure 3.** Summary of pre- and post-pandemic fare policy by WMATA and DASH.

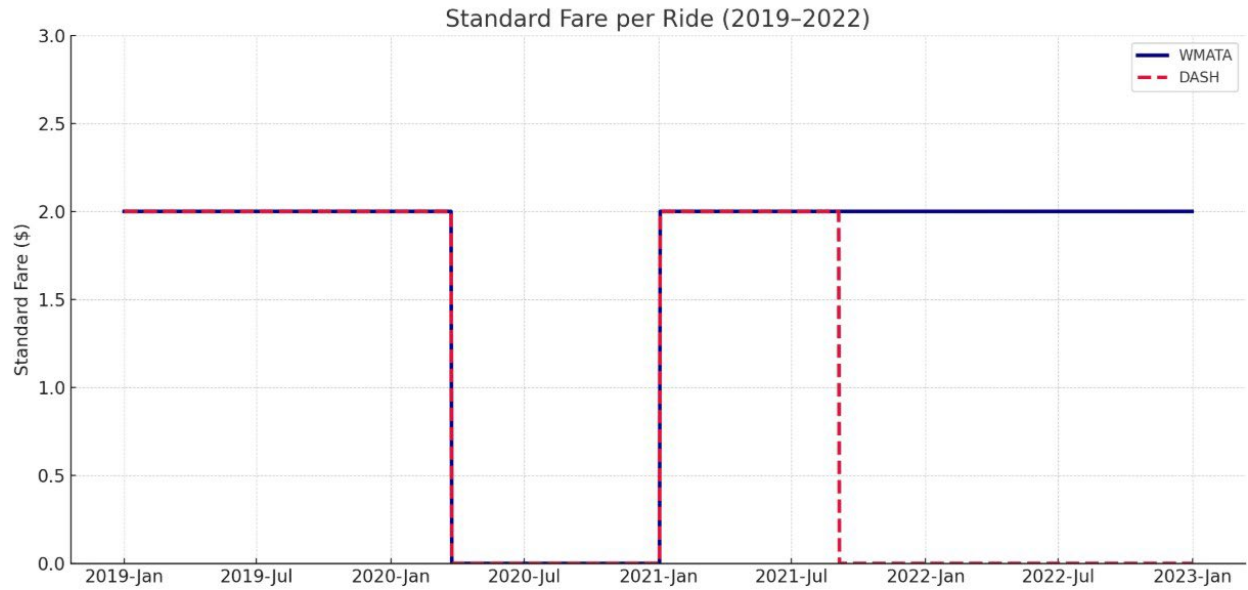

Our survey respondents were asked to recollect travel in August 2021 (before the FFTP started) and the time of the survey (June/July 2022), when the FFTP had been in place for already 9 months or more.

**Figure 4.** Monthly average of the hourly ozone measurements in Alexandria and in the control locations.

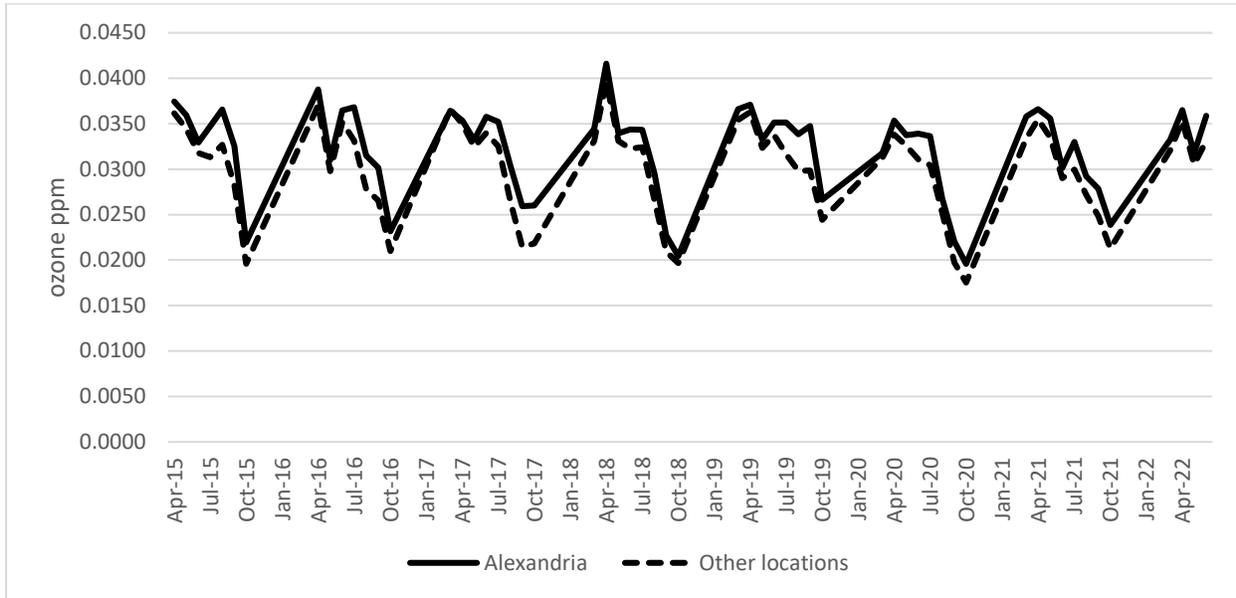

**Figure 5.** own survey: Car miles reduced per week since the free bus program/if the bus was free.

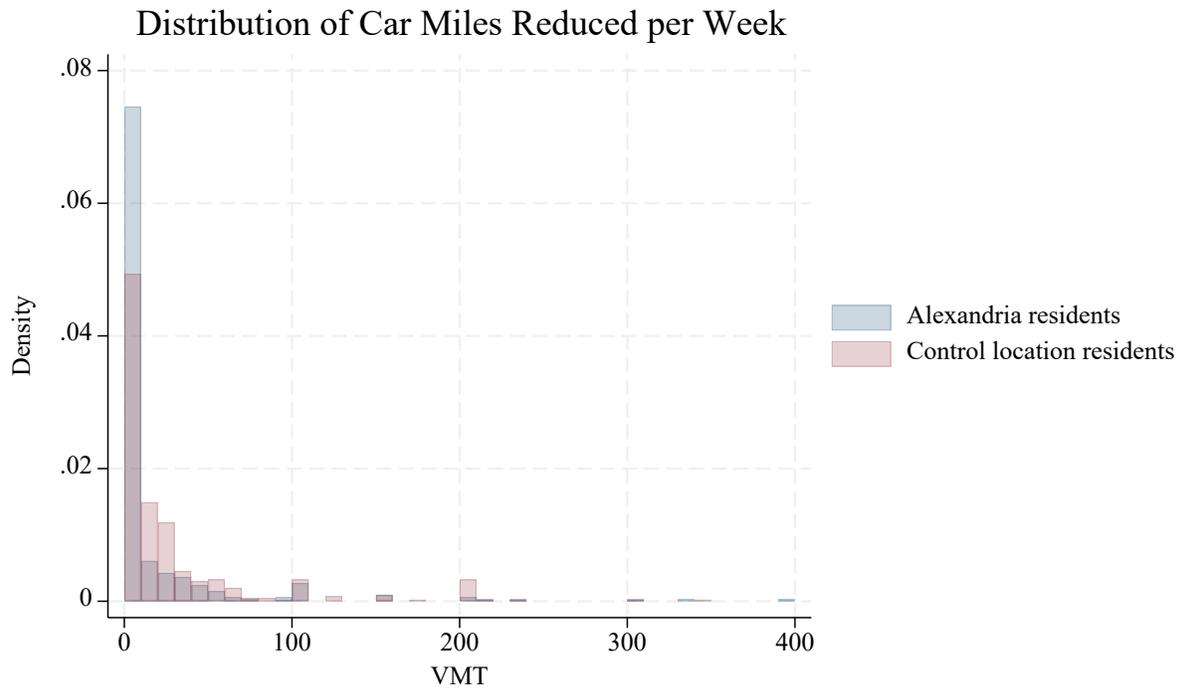

**Appendix.**

**Figure A.1.** Overlapping or partially overlapping Metrobus routes in Alexandria.

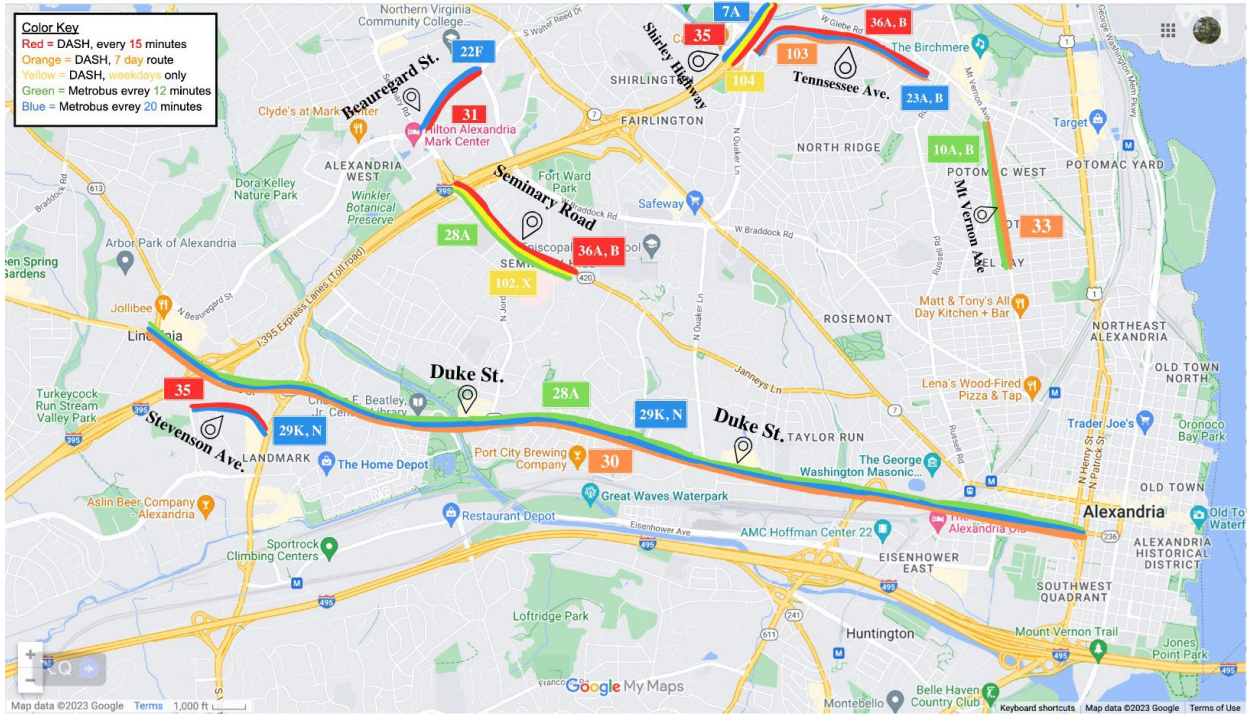

Map created by Pragna Yalamanchili.

**Figure A.2.** Ridership (number of daily boardings) on Metrobus routes in Alexandria that (partly) overlap with DASH bus routes.

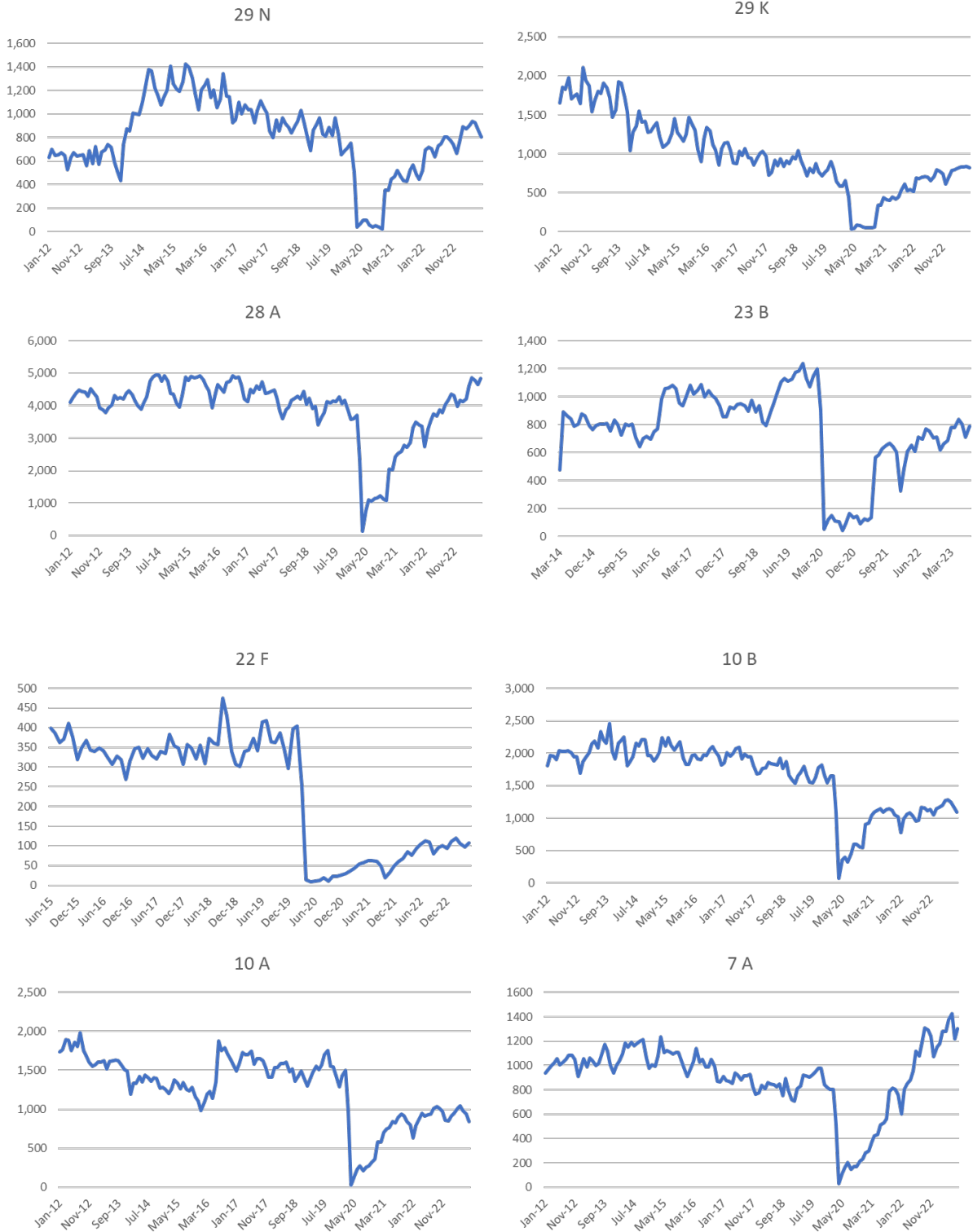

**Table A.1.** Composition of the sample by location. Alexandria = Treatment location; all other locations except Frederick: Control locations.

| Location | N | % |
|---|---|---|
| Alexandria | 338 | 33.9 |
| Arlington County | 45 | 4.51 |
| Calvert County | 19 | 1.91 |
| Charles County | 20 | 2.01 |
| City of Falls Church | 9 | 0.9 |
| Fairfax County | 109 | 10.93 |
| Frederick County | 44 | 4.41 |
| Manassas | 12 | 1.2 |
| Manassas Park | 4 | 0.4 |
| Montgomery County | 134 | 13.44 |
| Prince George's County | 85 | 8.53 |
| Prince William County | 58 | 5.82 |
| Washington DC | 120 | 12.04 |

**Table A.2.** Composition of the sample by employment status. Percent of the sample.

| Location | FT or PT employed | homemaker, students, or retired | other | total |
|---|---|---|---|---|
| Alexandria | 88.17 | 11.54 | 0.30 | 100 |
| Control locations | 66.02 | 29.59 | 4.39 | 100 |

**Table A.3.** Share of the sample by educational attainment.

| Educational level | Control locations | Alexandria | t test |
|---|---|---|---|
| Less than high school | 0.0179 | 0.003 | 0.22 |
| High school graduate | 0,1724 | 0.1686 | 0.06 |
| Some college | 0.2016 | 0.1716 | 0.50 |
| 2-.year degree | 0.0829 | 0.0976 | -0.23 |
| 4-year degree | 0.2618 | 0.2544 | 0.13 |
| Post-graduate education | 0.2098 | 0.2337 | -0.40 |
| Professional degree | 0.0537 | 0.0710 | -0.26 |

**Table A.4.** Distribution of the sample into the area's household income quartiles.

| Income quartile | Control locations | Alexandria |
|---|---|---|
| 1 | 24.55% | 25.74% |
| 2 | 24.23% | 26.92% |
| 3 | 26.83% | 21.89% |
| 4 | 24.39% | 25.44% |

**Table A.5.** Composition of the sample by race. Percent of the sample.

| Race | Control locations | Alexandria | t test |
|---|---|---|---|
| American Indian and Native Alaskan | 1.67 | 4.14 | -0.37547 |
| Asian | 7.89 | 7.1 | 0.122693 |
| Black/African American | 21.7 | 19.53 | 0.363215 |
| Multiracial | 1.52 | 1.18 | 0.051187 |
| Native Hawaiian and Pacific Islander | 0.76 | 1.78 | -0.15338 |
| Other | 3.79 | 5.62 | -0.2807 |
| White/Caucasian | 62.67 | 60.65 | 0.485565 |

**Table A.6.** Probit model. Dep. Var.: Usebus=using the bus with any frequency at time of survey. Standard errors in parentheses. N=997.

| | Coefficient |
|---|---|
| const | -0.0684 |
| | (0.1547) |
| Alexandria | 0.5199*** |
| | (0.0897) |
| Income quartile 2 | -0.2086* |
| | (0.1179) |
| Income quartile 3 | -0.0858 |
| | (0.1231) |
| Income quartile 4 | 0.1079 |
| | (0.1261) |
| College | -0.1362 |
| | (0.1043) |
| Postgrad | 0.0276 |
| | (0.1086) |
| Work | 0.3252*** |
| | (0.0993) |
| Has a car | -0.3093** |
| | -0.1452 |
| Commute < 10 miles | 0.0519 |
| | (0.0863) |

**Table A.7.** Own survey of DC metro area residents. Linear probability models of various dependent variables. OLS estimates; robust standard errors in parentheses.

|  | (1) morethan1 | (2) morethan2 | (3) morethan3 | (4) morethan4 |
|---|---|---|---|---|
| Cons | 0.0848* | 0.0737 | 0.0628 | 0.703*** |
|  | (0.0334) | (0.0419) | (0.0562) | (0.0678) |
| Alexandria | 0.0440 | 0.0577 | 0.0284 | -0.0208 |
|  | (0.0285) | (0.0433) | (0.0386) | (0.0428) |
| Work | 0.0662* | 0.0201 | 0.133** | 0.136* |
|  | (0.0274) | (0.0355) | (0.0451) | (0.0619) |
| Household income (thou. $) | 0.000430* | 0.001000*** | -0.000110 | -0.0000749 |
|  | (0.000212) | (0.000278) | (0.000308) | (0.000314) |
| College | -0.0313 | -0.133*** | 0.0832 | 0.0429 |
|  | (0.0312) | (0.0373) | (0.0509) | (0.0534) |
| Postgrad | -0.0408 | -0.0818 | 0.00844 | 0.0162 |
|  | (0.0332) | (0.0422) | (0.0507) | (0.0560) |
| Hispanic | 0.0339 | 0.0525 | 0.0250 | -0.0320 |
|  | (0.0366) | (0.0546) | (0.0487) | (0.0526) |
| African American | -0.00107 | -0.0181 | 0.0120 | -0.0544 |
|  | (0.0299) | (0.0392) | (0.0458) | (0.0516) |
| Commute is <10 miles | -0.00772 | 0.0151 | -0.0254 | -0.0826 |
|  | (0.0262) | (0.0354) | (0.0385) | (0.0426) |
| Annual miles driven (recoded) | -0.000652 | -0.00000111 | -0.000933* | 0.000311 |
|  | (0.000338) | (0.00106) | (0.000423) | (0.000327) |
| Annual miles missing | 0.0239 | 0.00718 | 0.0428 | -0.0552 |
|  | (0.0267) | (0.0354) | (0.0419) | (0.0450) |
| N | 997 | 530 | 467 | 467 |

*: significant at the 10% level; **: significant at the 5% level; ***: significant at the 1% level.